\documentclass[12pt, a4paper]{article}
\usepackage{cite}
\usepackage{amsmath,amssymb}
\input{colordvi.tex}
\usepackage{pifont}
\usepackage{comment}
\usepackage{bm}
\usepackage{url}
\usepackage{braket}

\usepackage[linktocpage]{hyperref}
\usepackage{ytableau}
\everymath{\displaystyle}
\usepackage{cancel}
\usepackage{mathtools}

\usepackage{ifpdf}
\ifpdf
  \usepackage{graphicx, hyperref, xcolor}     
\else     
  \usepackage[dvipdfmx]{graphicx, hyperref, xcolor}     
 \fi

\newcommand{\be}{\begin{equation}}
\newcommand{\ee}{\end{equation}}
\newcommand{\ba}{\begin{eqnarray}}
\newcommand{\ea}{\end{eqnarray}}

\definecolor{green}{rgb}{0, 0.65, 0.4}
\hypersetup{
setpagesize=false,
bookmarksnumbered=true,%
bookmarksopen=true,%
colorlinks=true,%
linkcolor=green,
urlcolor=green,
citecolor=green,
}

\DeclareUnicodeCharacter{2212}{-}
\begin{document}

\begin{titlepage}

\begin{center}

\hfill CERN-TH-2022-002\\

\vskip .75in

{\Large \bf
Discrete $R$-symmetry, Various Energy Scales and Gravitational Waves}

\vskip .75in

{\large
Gongjun Choi$^{(a)}$, Weikang Lin$^{(b)}$ and Tsutomu T. Yanagida$^{(b,c)}$
}
\vskip 0.25in

$^{(a)}${\em CERN, Theoretical Physics Department, Geneva, Switzerland}\\[.3em]
$^{(b)}${\em Tsung-Dao Lee 
Institute, Shanghai Jiao Tong University, Shanghai 200240, China}\\[.3em]
$^{(c)}${\em Kavli IPMU (WPI), The University of Tokyo,  Kashiwa, Chiba 277-8583, Japan}
\vskip 0.20in

\end{center}
\vskip .5in

\begin{abstract}

We present a supersymmetric model where energy scales of a discrete $R$-symmetry breaking ($Z_{6R}$) and cosmic inflation are commonly attributed to the confinement scale of a hidden $Sp(2)$ strong dynamics. Apart from these, SUSY-breaking scale, the Higgsino mass and the right-handed neutrino masses are all shown to stem from $Z_{6R}$ breaking scale inferred from CMB observables. We will show that the model is characterized by the SUSY-breaking soft mass $m_{\rm soft}\simeq100-1000{\rm TeV}$ and the reheating temperature $T_{\rm rh}\simeq10^{9}{\rm GeV}$. Then we discuss how these predictions of the model can be tested with the help of the spectrum of the gravitational wave induced by the short-lived cosmic string present during the reheating era.
\end{abstract}

\end{titlepage}


\renewcommand{\thepage}{\arabic{page}}
\setcounter{page}{1}
\renewcommand{\thefootnote}{\ding{83}\arabic{footnote}}
\setcounter{footnote}{0}
\renewcommand{\theequation}{\thesection.\arabic{equation}}


\newpage

\tableofcontents

\newpage

\section{Introduction}
\label{sec:Intro}
\setcounter{equation}{0}

When the Standard model (SM) is extended by the local supersymmetry (SUSY), it is believed that SUSY is broken at an energy scale higher than the electroweak (EW) scale due to the null observation of any sparticles in the LHC. Now that the observed vanishingly small cosmological constant results from the balance between a SUSY-breaking scale and a $R$-symmetry breaking scale in supergravity (SUGRA), $R$-symmetry should have been spontaneously broken at a certain time in the history of the universe at least prior to the EW phase transition era. 

As such, $R$-symmetry has been subject to questions about its nature. These include whether the symmetry is local or global and continuous or discrete. A global symmetry is argued to be easily broken by quantum gravity effects so that it is difficult to be exact~\cite{Banks:2010zn}. When applied to $R$-symmetry, the argument makes it difficult to discuss an $R$-symmetry breaking scale because the theory cannot control $R$-charged nonrenormalizable operators. Taking the attitude that conspiracy among fine-tuned coefficients of $R$-charged nonrenormalizable operators is never the decisive factor for determining an $R$-symmetry breaking scale, we focus our attention to a gauged $R$-symmetry. But it is very difficult to realize an anomaly free gauged $U(1)_{R}$ symmetry in the minimal SUSY SM (MSSM). Following this logic, we find that it is more probable to have the effective theory respecting a gauged discrete $R$-symmetry prior to generation of a non-zero constant term in the superpotential~\cite{Kurosawa:2001iq,Dine:2009swa}. 

On top of this, from the model building point of view, there are several merits to consider gauged discrete R-symmetries ($Z_{NR}$ with $N\in\mathbb{Z}$ and $N>2$). For some choices of $N$, the mixed anomalies of $Z_{NR}\otimes[G_{\rm SM}]^{2}$ vanish within MSSM where $G_{\rm SM}$ is a non-Abelian gauge group in the SM~\cite{Evans:2011mf}. For an anomaly free choice of $N$, the $R$-charge of the operator $H_{u}H_{d}$ becomes 4 modulo $N$, which prevents the Planck scale Higgsino mass~\cite{Harigaya:2013vja}.\footnote{We note that there can be still criticisms for taking the anomaly free conditions of discrete symmetries as one of guiding principles in low energy physics model buildings~\cite{Dine:2012mf}.} This fact, when combined with the requirement that $R$-charges of Yukawa coupling operators in the SM be 2 modulo $N$, naturally suppresses the dangerous proton decay operator {\bf 10}\,{\bf 10}\,{\bf 10}\,${\bf \bar{5}}$~\cite{Sakai:1981pk,Weinberg:1981wj}.\footnote{Here for notational convenience, we borrow representation notations {\bf 10} and ${\bf \bar{5}}$ of $SU(5)_{\rm GUT}$ to denote quarks and leptons in the SM.}

The appealing idea of having a gauged discrete $R$-symmetry in the theory, however, finds a dangerous cosmological problem when the symmetry breaking happened after the end of inflation~\cite{Zeldovich:1974uw,Kibble:1976sj}: It is unavoidable that domain walls form on the spontaneous breaking of $Z_{NR}$. This causes the universe to be quickly dominated by the domain wall and thus to be overclosed unless the symmetry breaking took place before the end of inflation. One may wonder whether the possibility of the breaking after the inflation can be saved with a sufficiently small explicit $R$-symmetry breaking term in the superpotential~\cite{Dine:2010eb}. Once an anomaly free $Z_{NR}$ is gauged, however, it does not admit such a possibility. Therefore, as a resolution to the domain wall problem, requiring the symmetry breaking to take place prior to the end of inflation can provide us with a lower bound on $R$-symmetry breaking scale in terms of either a Hubble expansion rate during inflation or a reheating temperature. The simplest solution of the domain wall problem is to consider the situation where the $R$ symmetry breaking dynamics drives the inflation at the same time.

On the other hand, $R$-symmetry is somewhat similar to spacetime symmetries in that it should be respected by every operator appearing in a superpotential. For the energy scale below the spontaneous $Z_{NR}$ breaking, this observation may arouse an interesting question whether dimensionful parameters of operators in the superpotential can be universally explained by powers of $Z_{NR}$ breaking scales. In light of this question, if a $Z_{NR}$ breaking scale could be related to an inflation scale, we can dream of the fascinating scenario where energy scales of $R$-symmetry breaking, inflation, SUSY-breaking and several dimensionful parameters in the MSSM share the common origin.

Motivated by the aforementioned questions, in this paper, we present a model where the listed various energy scales can be explained by a $Z_{6R}$ breaking scale (Sec.~\ref{sec:model}). Our choice for $Z_{6R}$ is based on the fact that it is the unique anomaly free discrete R-symmetry in the three-family MSSM~\cite{Evans:2011mf}. Considering the case in which the spontaneous $Z_{6R}$ breaking generates an inflaton potential and thus becomes connected to the inflation scale, we infer the $Z_{6R}$ breaking scale from the inflation dynamics consistent with cosmic microwave background (CMB) observables (Sec.~\ref{sec:inflation}). Then we further show how the infamous $\mu$-parameter (Higgsino mass) and the right handed neutrino mass can be connected to and explained by the $Z_{6R}$ breaking scale (Sec.~\ref{sec:various}). We shall also discuss the model's prediction on the reheating temperature and SUSY particle mass spectrum (Sec.~\ref{sec:reheating}), which can be possibly tested by the spectrum of the gravitational wave caused short-lived cosmic strings (Sec.~\ref{sec:GW}). From here on, we will use the same notation for a chiral superfield and its scalar component. Context discussed shall clarify which one is meant.

\section{Model}
\label{sec:model}
\setcounter{equation}{0}
In this section, we specify ingredients of our model by specifying the symmetry group and particle contents. In addition, we discuss how the spontaneous breaking of $R$-symmetry is realized in the model with the help of the hidden strong dynamics of $Sp(2)$. From here on, we take the Planck unit where the reduced Planck scale is set to the unity, i.e. $M_{P}=(8\pi G)^{-1/2}=1$.

In addition to the SM gauge group, the symmetry group the model assumes is given by 
\be
 G=\underbrace{Sp(2)\otimes Z_{6R}}_{{\rm gauge}}\otimes\underbrace{U(1)_{\Phi}\otimes Z_{4}}_{\rm global}\,.
\label{eq:symmetry}
\ee
As will be shown, the strong dynamics of $Sp(2)$ induces the spontaneous breaking of $Z_{6R}$ to $Z_{2R}$ as the theory enters the confined phase.\footnote{This way of inducing the spontaneous $Z_{6R}$ breaking is similar to the dynamical SUSY breaking based on Izawa-Yanagida-Intriligator -Thomas (IYIT) mechanism~\cite{Izawa:1996pk,Intriligator:1996pu}.} The particle contents and the charge assignment on them are shown in Table.~\ref{table1}. Concerning a gauged discrete $R$-symmetry the model obeys, we choose $Z_{6R}$ in accordance with the merits pointed out in Sec.~\ref{sec:Intro}. $Z_{4}$ can be regarded as a subgroup of the global $U(1)_{\rm B-L}$ symmetry where B (L) stands for the baryon (lepton) number.

In general, for $Sp(N)$ supersymmetric gauge theory with $N_{F}=2(1+N)$ chiral superfields $Q_{i}$ ($i=1-N_{F}$) transforming as the fundamental representation, the mixed anomaly of $Z_{6R}\otimes Sp(N)^{2}$ vanishes when the following condition is satisfied~\cite{Ibanez:1991hv,Ibanez:1991pr,Ibanez:1992ji}
\be
3\times R[\lambda_{a}]+\frac{1}{2}\times\left\{\sum_{i=1}^{N_{F}}(R[Q_{i}]-1)\right\}=\frac{6}{2}\times\ell\qquad(\ell\in\mathbb{Z})\,,
\label{eq:Rcharge1}
\ee
where $R[\lambda_{a}]=1$ and $R[Q_{i}]$ are $R$-charges of the gaugino and $Q_{i}$. Particularly for $N=2$ ($N_{F}=6$), we see that Eq.~(\ref{eq:Rcharge1}) holds true as long as $R[Q_{i}]$ is an integer. This explains our choice for $Sp(2)$ as a gauge group for the hidden strong dynamics. For an energy scale below the dynamical scale ($\Lambda_{*}$) of $Sp(2)$, the theory is known to be described by $(2N+1)(N+1)$ composite meson fields $\mathcal{M}_{ij}\equiv(4\pi)\langle Q_{i}Q_{j}\rangle/\Lambda_{*}$ with the deformed moduli constraint ${\rm Pf}(\mathcal{M}_{ij})=\Lambda_{*}^{3}$~\cite{Seiberg:1994bz}. 

\begin{table}[t]
\centering
\begin{tabular}{|c||c|c|c|c|c|c|c|c|c||c|c|} \hline
 & $\Phi$ & $Q_{i}$ & $S_{ij}$& $H_{u}$ & $H_{d}$ & $N$ & $\lambda_{ij}$ & $\delta_{\Phi}$\\
\hline
$Sp(2)$ &  -  &  $\ytableausetup{textmode, centertableaux, boxsize=0.6em}
\begin{ytableau}
 \\
\end{ytableau}$  & - & - & - &- & - & -\\
\hline
$Z_{6R}$ &  +1  &  +1  & 0 & $x$ & $4-x$ &0 & 0 & 0\\
\hline
$U(1)_{\Phi}$ &  -1  &  0  & 0 & 0 & 0 &0 & +2& +2\\
\hline
$Z_{4}$ &  +1  &  +1  & +2 & +2 & +2 &+1 & 0& +2\\
\hline
\end{tabular}
\caption{Charge assignment of chiral superfields in the model under the gauge group in Eq.~(\ref{eq:symmetry}) and the global $U(1)_{\Phi}$ for the phase rotation of $\Phi$. $\Phi$ is the inflaton chiral multiplet, $H_{u}$ $(H_{d})$ the up (down)-type Higgs chiral multiplet in the MSSM and $N$ the right-handed neutrino chiral multiplet. The parameters of the model $\lambda_{ij}$ and $\delta_{\Phi}$ are regarded as spurions.} 
\label{table1} 
\end{table}

At a high energy scale at which $Sp(2)$ is in the perturbative regime, the part of the superpotential of the theory reads
\be
W_{\rm total}\ni W_{\cancel{R}}+W_{HN}\,.
\label{eq:superpotential}
\ee
In Eq.~(\ref{eq:superpotential}),  $W_{\cancel{R}}$ is the part of $W_{\rm total}$ responsible for $R$-symmetry breaking and the inflation
\be
W_{\cancel{R}}=-a_{ij}Q_{i}Q_{j}S_{ij}+\lambda_{ij}S_{ij}\Phi\Phi+\delta_{\Phi}\Phi\Phi\,, 
\label{eq:WR}
\ee
where $a_{ij}$ and $\lambda_{ij}$ are dimensionless coupling constants and the sum over the repeated indices is assumed implicitly. 

On the other hand, $W_{HN}$ contains the mass terms for the Higgsino and the right-handed neutrinos\footnote{There are more operators contributing to $W_{HN}$ which are of the form $\sim\Phi^{4}H_{u}H_{d}$ and $\sim\Phi^{2}NN$ respecting $Z_{6R}$. As discussed later, when these are accompanied by the spurion fields $\lambda_{ij}$ and $\delta_{\Phi}$, their contribution to the Higgsino and right handed neutrino mass is comparable to operators in Eq.~(\ref{eq:WHN}). Thus for our purpose, Eq.~(\ref{eq:WHN}) suffices.}
\be
W_{HN}=b_{ij}(Q_{i}Q_{j})^{2}H_{u}H_{d}+c_{ij}Q_{i}Q_{j}NN\,,
\label{eq:WHN}
\ee
where $b_{ij}$ and $c_{ij}$ are a dimensionless coupling constant. We implicitly assumed three species of the right handed neutrinos for which there exist three different $c_{ij}$s.

Without loss of generality, we can choose the moduli space in the confined phase of $Sp(2)$ such that vacuum expectation values (VEVs) of $Q_{i}$ fields satisfy 
\ba
&&\langle Q_{1}Q_{2}\rangle=\langle Q_{3}Q_{4}\rangle=\langle Q_{5}Q_{6}\rangle=v^{2}=\frac{\Lambda_{*}^{2}}{4\pi}\cr\cr
&&\langle Q_{i}Q_{j}\rangle=0\quad {\rm for}\quad j-i\neq1
\,,
\label{eq:vev}
\ea
where $\Lambda_{*}$ is the dynamical scale of $Sp(2)$. From Eq.~(\ref{eq:vev}), it becomes self-evident that the spontaneous breaking of $Z_{6R}$ to $Z_{2R}$ occurs when $Sp(2)$ becomes strongly coupled. In order to simplify the first term in Eq.~(\ref{eq:WR}), we make the following definition of the chiral superfield $S$ 
\be
S\equiv a_{12}S_{12}+a_{34}S_{34}+a_{56}S_{56}\,.
\label{eq:S}
\ee
Then in the confined phase, we can rewrite Eq.~(\ref{eq:WR}) as
\be
W_{\cancel{R}}=-v^{2}S+\lambda S\Phi\Phi+\delta_{\Phi}\Phi\Phi\,,
\label{eq:WR2}
\ee
where we assumed $\sum_{ij}\lambda_{ij}/a_{ij}=\lambda$ for the simplicity of the analysis. 

The superpotential in Eq.~(\ref{eq:WR2}) provides the scalar potential for $S$ and $\Phi$ with account taken of the K\"{a}hler potential below,
\be
K(\Phi,S)=|S|^{2}+|\Phi|^{2}+c_{1}|S|^{2}|\Phi|^{2}+...\,,
\label{eq:kahler}
\ee
where the ellipsis stands for higher powers of $|S|$ and $|\Phi|$. From here on, we assume the suppression of the higher order terms in $K(\Phi,S)$ above so that the three terms in Eq.~(\ref{eq:kahler}) are dominant.\footnote{Of course, this assumption is hardly justified in the model we present in this section since much more higher dimensional operators including higher powers of $|\Phi|$ and $|S|$ can be allowed by the symmetry of the model and thus expected to be present. Nevertheless, for our purpose of relating R-symmetry breaking scale and the inflation scale, we rely on this assumption for suppression and make the prediction of the model for the CMB observables in accordance with the assumption.}  

We end this section by commenting on the values of $\lambda$ and $\delta_{\Phi}$. As can be seen in the next section, we shall consider the case where the last term in Eq.~(\ref{eq:WR2}) is irrelevant for the inflationary dynamics. Namely the last term's contribution to the inflaton potential during inflation is sub-dominant. However, on acquisition of VEV of $\Phi$ at the end of inflation, the last term generates the constant term for the superpotential. Given $\langle\Phi\rangle\sim2$ at the end of inflation, we will set $\delta_{\Phi}$ to be of order a gravitino mass ($\sim v^{4}$) in order to correctly produce the vanishingly small cosmological constant. In addition, $\langle\Phi\rangle\sim2$ at the end of inflation will further require $\lambda\sim v^{2}\simeq10^{-6}$. We take the attitude to treat $\delta_{\Phi}$ and $\lambda$ as spurions so that their smallness is originated from the breaking of the global symmetries $U(1)_{\Phi}$ and $Z_{4}$ shown in Table.~\ref{table1}.

\section{Inflationary Dynamics}
\label{sec:inflation}
\setcounter{equation}{0}
In this section, we discuss how the dynamically generated superpotential in Eq.~(\ref{eq:WR2}) and the K\"{a}hler potential in Eq.~(\ref{eq:kahler}) can lead to the inflationary expansion of the early universe~\cite{Starobinsky:1980te,Guth:1980zm,Linde:1981mu,Albrecht:1982wi} for a certain range of parameters $\lambda$ and $v$. For the pre-inflationary era, we envision the situation where the universe is dominated by the thermal bath since the Planck time $t\sim M_{P}^{-1}$. Starting from $T\sim M_{P}$, the temperature of the universe continues to decrease in the expanding background until the inflationary era is reached. $\Phi$ multiplet remains decoupled from the thermal bath because of the smallness assumed for $\lambda\sim v^{2}\simeq10^{-6}$ while $Q$, $S$ and the gluon multiplet of $Sp(2)$ are in thermal equilibrium and thus $\langle S\rangle=\langle Q\rangle\sim0$. When $T\simeq v$ is met, $Sp(2)$ theory becomes strongly coupled and $Z_{6R}$ is broken down to $Z_{2R}$. Then the domain wall associated with the discrete $R$-symmetry breaking forms. Later when the inflaton potential energy dominates the energy at a certain spatial region, the single field slow-roll inflation gets started. 

At the Planck time, in principle $\Phi$ can be any value within $[-M_{P},M_{P}]$  because the field fluctuation is comparable to the Hubble expansion rate prior to the inflation ($H_{\rm pre}$), i.e. $\delta\Phi\simeq H_{\rm pre}/(2\pi)\sim M_{P}$. Now we can readily expect the presence of the spatial region containing two points $x_{+}$ and $x_{-}$ separated by more than $1/H_{\rm pre}$ with $\Phi(x_{+})\simeq M_{P}$ and $\Phi(x_{-})\simeq -M_{P}$. With $\Phi$ chosen monotonic within $[x_{-},x_{+}]$, its analyticity and continuity guarantee that there exists a spatial point $x_{0}$ corresponding to the zero field value ($\Phi(x_{0})=0$) inside the Hubble patch of the radius $1/H_{\rm pre}$. Let us denote this kind of Hubble patch as $H_{x_{0}}$. As far as the Hubble patch $H_{x_{0}}$ is concerned, we may regard the dynamics of $Q$ as unaffected by $\Phi$ thanks to $|\Phi|<\!\!<1$ and thus naturally expect $\delta Q\lesssim H_{\rm pre}/(2\pi)$.\footnote{For the region where $|\Phi|<\!\!<1$, both $Q$ and $S$ have convex potentials centered on the origin of the field space. So their fluctuations are at most that of a massless scalar.} Given that the curvature of $\Phi$ field potential amounts to $\sim\lambda Q^{2}$ which is at most $\lambda H_{\rm pre}^{2}$, the smallness of $\lambda$ implies $\dot{\Phi}\approx0$ especially for the spatial region near $x_{0}$. This point assures us that the presence of $x_{0}$ within the Hubble patch $H_{x_{0}}$ persists as the universe cools down prior to the inflation.

When the universe enters the confined phase of $Sp(2)$ (still in the pre-inflationary era), the superpotential of $\Phi$ within the Hubble patch $H_{x_{0}}$ effectively becomes of the form Eq.~(\ref{eq:WR2}) and we expect the domain wall associated with $Z_{6R}$ breaking to form around $H_{x_{0}}$.\footnote{For the case where the separation between the two points $x_{\pm}$ is smaller than $1/H_{\rm pre}$, a value of $Q$ varies significantly within $H_{x_{0}}$. So the domain wall associated with $Z_{6R}$ breaking may form within a Hubble patch of the radius $1/H_{\rm pre}$. This makes the evolution of $\Phi$ in $H_{x_{0}}$ unclear and complicated. It might be still probable to have the inflation in such a patch, but more rigorous exploration dealing with the coupled system of $Q$ and $\Phi$ is needed for further discussion. Thus we focus on the opposite situation as specified in the main text. } In SUGRA, the scalar potential is given by 
\be
V=e^{K}\left[\sum_{m,n}\left(\frac{\partial^{2}K}{\partial X_{m}\partial X_{n}^{*}}\right)^{-1}D_{X_{m}}WD_{X_{n}^{*}}W^{*}-3|W|^{2}\right]\,,
\label{eq:VSUGRA}
\ee
where $X_{m}$ is a chiral superfield, the subscript on $X_{m}$ distinguishes different chiral superfields and $D_{X_{m}}W=(\partial W/\partial X_{m})+W(\partial K/\partial X_{m})$ was defined. We assume a set of coefficients of operators forming the potential of $S$ so as to have $S=0$.\footnote{$S=0$ can be easily justified even by relying on the positive Hubble induced mass that drives evolution of $S$ towards the origin of the field space.} By substituting $W_{\cancel{R}}$ in Eq.~(\ref{eq:WR2}) and $K(\Phi,S)$ in Eq.~(\ref{eq:kahler}) into Eq.~(\ref{eq:VSUGRA}), we obtain the following potential for $\Phi$ at $S=0$
\be
V(\Phi)\simeq e^{|\Phi|^{2}}\left[\frac{|v^{2}-\lambda\Phi\Phi|^{2}}{(1+c_{1}\Phi\Phi)}+\delta_{\Phi}^{2}|\Phi|^{2}\right]
\,.
\label{eq:Vscalar}
\ee
Identifying the real part of $\Phi$ with the inflaton field, i.e. $\phi=\Phi/\sqrt{2}$ and ignoring the subdominant term, we obtain the following potential for $\phi$ relevant for the slow-rolling, 
\be
V(\phi)\simeq v^{4}e^{\frac{\phi^{2}}{2}}\left(1-\kappa\frac{\phi^{2}}{2}\right)^{2}\left(1+c_{1}\frac{\phi^{2}}{2}\right)^{-1}\,,
\label{eq:VPhi}
\ee
where we defined $\kappa\equiv\lambda/v^{2}$.

On formation of $V(\phi)$ in Eq.~(\ref{eq:VPhi}), we expect that the single-field slow-roll inflation within $H_{x_{0}}$ gets started with the initial inflaton field value close to zero. We have seen that the presence of $x_{0}$ in $H_{x_{0}}$ can be ensured and thus the inflation can occur in $H_{x_{0}}$ with $V(\phi)$ just like the topological inflation~\cite{Vilenkin:1994pv}. The degree of slow-rolling is measured by the two parameters $\epsilon$ and $\eta$
\be
\epsilon(\phi;c_{1},\kappa)\equiv\frac{1}{2}\left(\frac{V^{'}}{V}\right)^{2}\quad,\quad\eta(\phi;c_{1},\kappa)\equiv\left(\frac{V^{''}}{V}\right)\,,
\label{eq:slowroll}
\ee
which are functions of $\phi$ and depend on $c_{1}$ and $\kappa$. Let us use the subscript ``$\star$" (``end") for quantities evaluated at the time of the horizon exit of the CMB pivot scale (at the end of inflation). We define $\phi_{\rm end}$ as a solution to the equation $\epsilon(c_{1},\kappa)=1$. 

For a given set of $(v,c_{1},\kappa)$ and $0<\phi_{\star}<1$, using Eq.~(\ref{eq:VPhi}) and Eq.~(\ref{eq:slowroll}), we can compute the prediction of the model for the following CMB observables including the spectral index ($n_{s}$) and the amplitude ($A_{s}$) of the power spectrum for the comoving curvature perturbation, and  the tensor to scalar ratio ($r$) at the CMB pivot scale $k_{\star}=0.05{\rm Mpc}^{-1}$
\be
A_{s}=\frac{1}{12\pi^{2}}\frac{V(\phi_{\star})^{3}}{V^{'}(\phi_{\star})^{2}},\quad n_{s}-1=-6\epsilon_{\star}+2\eta_{\star},\quad r=16\epsilon_{\star}\,,
\label{eq:ns}
\ee
and the number of e-foldings during the inflation
\be
N_{\star}=\int_{t_{\star}}^{t_{\rm end}}Hdt\simeq-\int_{\phi_{\star}}^{\phi_{\rm end}}\left(\frac{3H^{2}}{V^{'}}\right)d\phi\simeq-\int_{\phi_{\star}}^{\phi_{\rm end}}\frac{V}{V'}d\phi\,, 
\label{eq:N*}
\ee
where the slow-roll approximation $\dot{\phi}\simeq-V^{'}/3H$ and $3H^2\simeq V$ were used.

For these values, we adopt $A_{s}=2.1\times10^{-9}$, $n_{s}=0.9649\pm0.0042$ ($68\%$ C.L., Planck TT,TE,EE+lowE+lensing)~\cite{Planck:2018jri} and $r<0.036$ ($95\%$ C.L., BICEP/Keck)~\cite{BICEP:2021xfz}. In addition, we shall see that $48.5\lesssim N_{\star}\lesssim53$ for a consistent reheating scenario with $T_{\rm rh}\sim10^{9}{\rm GeV}$ and a reasonable reheating equation of state.

\begin{figure}
  \centering
  \includegraphics[width=0.95\hsize]{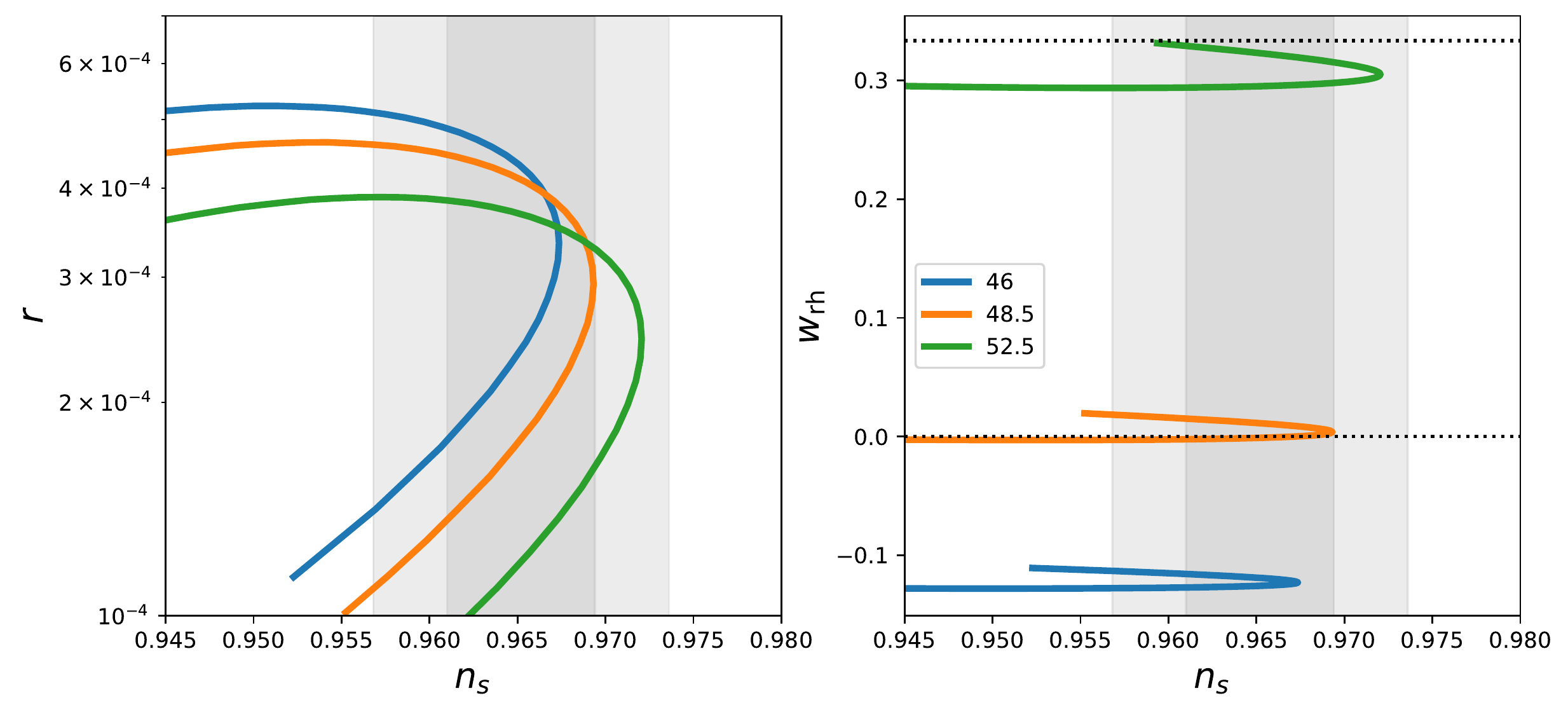}
  \caption{Some predicted quantities in our model taking $c_1=0.435$. Left: the prediction of the inflation model for the scalar spectral index ($n_{s}$) and the tensor-to-scalar ratio ($r$) for each of the specified $N_{\star}=46$ (blue), $48.5$ (orange) and $52.5$ (green). The grey vertical bands show the $1\sigma$ and $2\sigma$ constraints of $n_s$. Right: the inferred reheating equation of state ($w_{\rm rh}$) versus $n_{s}$ taking $T_{\rm rh}\simeq10^9{\rm{GeV}}$. When $0<w_{\rm rh}<1/3$ is considered, we can see that $48.5\lesssim N_*\lesssim53$.
  }
  \label{fig1}
\end{figure}

By scanning the parameter space of $(v,c_{1},\kappa)$ and $0<\phi_{\star}<1$, it is realized that $v\sim1.5\times10^{-3}$, $c_{1}\gtrsim0.4$, $\kappa\sim0.3$ and $\phi_{\star}\sim0.6$ achieve a good fit to the CMB observables. Accordingly, the inflation dynamics triggered by confined phase of $Sp(2)$ gauge theory results in the relatively low tensor-to-scalar ratio $r=\mathcal{O}(10^{-4})$. In Fig.~\ref{fig1}, we show the prediction of the inflation model for $n_{s}$ and the tensor-to-scalar ratio $r$ corresponding to some examples of $N_{\star}$.

\section{Various Energy Scales}
\label{sec:various}
In this section, we first infer the $R$-symmetry breaking scale, i.e. $v$ in Eq.~(\ref{eq:vev}), from CMB observables discussed in Sec.~\ref{sec:inflation}. Then, by demanding that the model be able to accommodate the successful electroweak symmetry breaking (EWSB) in the SUGRA framework, we infer Higgsino mass ($\mu_{H}$), gravitino mass ($m_{3/2}$) and thus SUSY-breaking scale ($M_{\cancel{\rm SUSY}}=\sqrt{|F|}$). We will see that our model unifying spontaneous $Z_{6R}$ breaking and inflationary dynamics is actually nothing but a way of realizing the early universe physics in the pure gravity mediation scenario~\cite{Ibe:2011aa}. As such, the model will be shown to be subject to the constraint on the reheating temperature not to have too much relic abundance of wino dark matter (DM) candidate. 

In our model, the discrete $R$-symmetry is taken to be the origin of a variety of energy scales. As such, its breaking scale is intended to account for the only dimensionful parameter in MSSM, i.e. Higgsino mass ($\mu_{H}$) along with the right-handed neutrino masses through Eq.~(\ref{eq:WHN}). In accordance with this feature of the model, now we are in the position to discuss these mass scales and resultant physics in light of the value $v\sim1.5\times10^{-3}$ obtained from CMB observables discussed in Sec.~\ref{sec:inflation}.

Above all, $H_{u}H_{d}$ having $R$-charge 4, it couples to four $Q$ chiral superfields. Thus the model predicts $\mu_{H}=\mathcal{O}(0.1)\times v^{4}\simeq100-1000{\rm TeV}$ at the tree level.\footnote{In Eq.~(\ref{eq:WHN}), we take $b_{ij}, c_{ij}=\mathcal{O}(0.1)$.} In addition, the Majorana mass terms for the right handed neutrinos couple to two $Q$ chiral superfields since they have zero $R$-charge. This provides the supersymmetric right handed neutrino mass $m_{N}=\mathcal{O}(0.1)\times v^{2}\simeq10^{11}-10^{12}{\rm GeV}$. Thus, the model can explain $m_{N}$ required for the leptogenesis successfully with the aid of the discrete $Z_{6R}$ symmetry.

\begin{table}[t]
\centering
\begin{tabular}{|c||c|} \hline
 Relevant physics & Energy scale \\
\hline\hline
$R$-symmetry breaking      &  $v$ ($\sim10^{15}{\rm GeV}$)    \\\hline
SUSY breaking  &  $v^{2}$ ($\sim10^{12}{\rm GeV}$)    \\\hline
Inflation scale ($H_{\rm inf}$) &  $v^{2}$ ($\sim10^{12}{\rm GeV}$)  \\\hline\hline
Higgsino mass ($\mu_{H}$)      &  $v^{4}$ ($\sim10^{5}-10^{6}{\rm GeV}$)    \\\hline
Right-handed neutrino mass  ($m_{N}$)    & $v^{2}$ ($\sim10^{11}-10^{12}{\rm GeV}$)   \\\hline
\end{tabular}
\caption{Various energy scales which are the direct consequence of $Z_{6R}$ symmetry. The energy scales are written in terms of the vev of $Q$ field, i.e. $v$. For the numbers of the energy scales for the dimensionful parameter, $\mathcal{O}(0.1)$ coupling constants are taken into account.}
\label{table:energyscales} 
\end{table}

Now given the value of Higgsino mass, it is realized that we can infer a gravitino mass $m_{3/2}$ from the two conditions for EWSB\footnote{Two conditions means the negative curvature of the Higgs potential at the origin of field space and the lower bounded potential.}
\be
(|\mu_{H}|^{2}+m_{H_{u}}^{2})(|\mu_{H}|^{2}+m_{H_{d}}^{2})\simeq(B\mu_{H})^{2}\,.
\label{eq:EWSB}
\ee
Because the scalar soft masses $m_{H_{u}}$, $m_{H_{d}}$ and $B$ are of the order $m_{3/2}$ due to the SUGRA effect~\cite{Nilles:1983ge}, $\mu_{H}=b_{ij}\langle Q_{i}Q_{j}\rangle^{2}=\mathcal{O}(m_{3/2})$ should be the case. Hence, EWSB and the universal scalar soft SUSY-breaking masses of the order $m_{3/2}$ at the tree level in SUGRA give the important information for the gravitino mass of the model, i.e. $m_{3/2}\simeq\mu_{H}\simeq100-1000{\rm TeV}$.

Once $m_{3/2}$ is known, now we can infer the SUSY-breaking scale based on the observation for the vanishingly small cosmological constant. From the leading order contribution to the potential in Eq.~(\ref{eq:VSUGRA}), the SUSY-breaking scale reads
\be
M_{\cancel{\rm SUSY}}^{2}=|F|\simeq \sqrt{3}m_{3/2}=\mathcal{O}(0.1)\times v^{4}
\ee
where $F$ is the auxiliary component of a SUSY-breaking field. As a summary of our discussion thus far, we show the various energy scales resulting from the structure of the model and CMB observables in Table.~\ref{table:energyscales}.

With soft SUSY-breaking scalar masses and $\mu_{H}$ all comparable to $m_{3/2}$, as for the mass spectrum of the model, the one remaining question regards the gaugino masses. Then without extending the model for the SUSY-breaking mediation to the visible sector, what would be the prediction for gaugino masses in the current minimal scenario? The gaugino mass can be generated through SUGRA effect at the one-loop level (a.k.a the anomaly mediation)~\cite{Giudice:1998xp,Randall:1998uk,Dine:1992yw}. This means the gaugino mass is given by $|M_{a}|\simeq b_{a}(g_{a}^{2}/(16\pi^{2}))m_{3/2}$ where the subscript $a$ is the group index, $g_{a}$ is the gauge coupling and $b_{a}$ is the beta function coefficient at the one loop level. 

 With $m_{3/2}=\mathcal{O}(0.1)\times v^{4}$, now it is realized that this set-up and mass spectrum are precisely what's envisioned in the pure gravity mediation scenario~\cite{Ibe:2011aa}. Wino becomes the lightest supersymmetric particle (LSP) thereof and thus DM candidate. For the wino mass $M_{2}\simeq2.7 {\rm TeV}$, the thermal relic of the wino ($\Omega^{\rm th}_{\rm wino}(M_{2})h^{2}$) can explain the current DM abundance~\cite{Hisano:2006nn,Cirelli:2007xd}. For a smaller $M_{2}$ leading to $\Omega^{\rm th}_{\rm wino}(M_{2})h^{2}<\!\!<0.12$, still the non-thermal production from the decay of the gravitino can give rise to $\Omega^{\rm nth}_{\rm wino}(M_{2},T_{\rm rh})h^{2}\simeq0.12$ that can explain the DM abundance today depending on $T_{\rm rh}$. $\Omega^{\rm nth}_{\rm wino}(M_{2},T_{\rm rh})h^{2}$ being proportional to $M_{2}T_{\rm rh}$, $\Omega^{\rm nth}_{\rm wino}(M_{2},T_{\rm rh})h^{2}\simeq0.1$ is satisfied for $M_{2}\simeq2{\rm TeV}$ and $T_{\rm rh}\simeq10^{9}{\rm GeV}$~\cite{Ibe:2011aa}.

Therefore, we come to see that viability for explaining the gaugino masses within the current minimal scenario depends on whether the model can be consistent with $\Omega_{\rm DM}h^{2}\simeq0.12$ with a proper choice of $T_{\rm rh}$. From the wino mass generated via the anomaly mediation, $M_{2}\simeq300{\rm GeV}-3{\rm TeV}$ is expected. Thus, the current abundance of the wino $\Omega_{\rm wino}h^{2}=\Omega^{\rm th}_{\rm wino}(M_{2})h^{2}+\Omega^{\rm nth}_{\rm wino}(M_{2},T_{\rm rh})h^{2}$ can avoid exceeding $\Omega_{\rm DM}h^{2}$ when $T_{\rm rh}\lesssim10^{9}{\rm GeV}$ is satisfied. 

In the naive estimate of $T_{\rm rh}$ using $\Gamma_{\phi}\simeq H(T_{\rm rh})$, $T_{\rm rh}\simeq\sqrt{m_{\Phi}}\simeq \sqrt{\kappa}v^{2}=\mathcal{O}(10^{12}){\rm GeV}$ up to $\mathcal{O}(0.1)$ coupling constant factor where $m_{\Phi}$ is the effective inflaton mass read from Eq.~(\ref{eq:VPhi}). Thus it is not clear whether our inflation model is able to accommodate $T_{\rm rh}$ as small as $10^{9}{\rm GeV}$. It is known that $T_{\rm rh}$ is closely related to the shape of an inflaton potential~\cite{Liddle-Leach:2003,Munoz:2014eqa,Cook:2015vqa}. In the next section, we shall address this question for $T_{\rm rh}$ to see whether the model can explain the gaugino mass based on the anomaly mediation without any further extension in the model.

\section{Reheating}
\label{sec:reheating}
In this section, we discuss the prediction of the model for the reheating temperature $T_{\rm rh}$ based on the slope and the curvature of the inflaton potential in Eq.~(\ref{eq:VPhi}). We denote the time average value of the equation of the state of the universe during the reheating stage by $w_{\rm rh}$. We expect $w_{\rm rh}$ to be positive and close to $0$ because the inflaton field went through the coherent oscillation after the inflation ends with the parabolic convex potential shape for $\phi>\phi_{\rm end}>1$ \cite{Turner:1983of}. We begin with the review of the procedure to compute $T_{\rm rh}$ based on \cite{Munoz:2014eqa,Cook:2015vqa}. For a given $w_{\rm rh}$, eventually we will see that $T_{\rm rh}$ is closely related to inflationary dynamics via a model's prediction discussed in Sec.~\ref{sec:inflation}.

\subsection{$T_{\rm rh}$ and $w_{\rm rh}$}
\label{sec:Trhandwrh}
\setcounter{equation}{0}
Let us denote the scale factor and the Hubble expansion rate at the horizon-exit of comoving wavenumber $k$ by $a_{k}$ and $H_{k}$ respectively. Then for the pivot scale $k_{\star}=0.05{\rm Mpc}^{-1}$, the ratio $k_{\star}/(a_{0}H_{0})$ can be written as
\be
\frac{k_{\star}}{a_{0}H_{0}}=\frac{a_{k_{\star}}H_{k_{\star}}}{a_{0}H_{0}}=\frac{a_{k_{\star}}}{a_{\rm end}}\frac{a_{\rm end}}{a_{\rm rh}}\frac{a_{\rm rh}}{a_{\rm 0}}\frac{H_{k_{\star}}}{H_{0}}\,, 
\label{eq:scaleratio}
\ee
where each subscript $0$, end, rh, eq, stands for the end of inflation, the end of reheating, the matter-radiation equality. By taking the logarithm for both sides, we can rewrite Eq.~(\ref{eq:scaleratio}) as
\be
\ln\left(\frac{k_{\star}}{a_{0}H_{0}}\right)=-N_{\star}-N_{\rm rh}-N_{\rm R0}-\ln\left(\frac{H_{0}}{H_{k_{\star}}}\right)\,,
\label{eq:scaleratio2}
\ee
where we used the parametrization $(a_{\rm end}/a_{k_{\star}})=e^{N_{\star}}$, $(a_{\rm rh}/a_{\rm end})=e^{N_{\rm rh}}$ and $(a_{0}/a_{\rm rh})=e^{N_{\rm R0}}$. Using the entropy conservation in the universe from the reheating until today, one can replace $-N_{\rm R0}$ on the right hand side with the expression including $T_{\rm rh}$. The aforesaid entropy conservation gives
\be
g_{*s,{\rm rh}}a_{\rm rh}^{3}T_{\rm rh}^{3}=\left(2+\frac{7}{8}\times2N_{\rm eff}\times\frac{4}{11}\right)\times a_{0}^{3}\times T_{0}^{3}\,,
\label{eq:entropyconservation}
\ee
where $g_{*s,{\rm rh}}$ is the effective number of degrees of freedom in entropy, $N_{\rm eff}$ is the effective number of neutrino species and $T_{0}$ is the current photon temperature. For neutrino temperature, we used $T_{\nu0}=(4/11)^{1/3}T_{0}$. By taking $N_{\rm eff}=3$ for simplicity, we obtain
\be
N_{\rm R0}=\ln\left(\frac{T_{\rm rh}}{T_{0}}\right)-\frac{1}{3}\ln\left(\frac{43}{11g_{*s,{\rm rh}}}\right)\,.
\label{eq:Nrd}
\ee

From the ratio of the energy density of the universe at reheating $\rho_{\rm rh}$ to that at the end of inflation $\rho_{\rm end}$, we have
\be
\frac{\rho_{\rm rh}}{\rho_{\rm end}}=\frac{\frac{\pi^{2}}{30}g_{*,{\rm rh}}T_{\rm rh}^{4}}{\frac{3}{2}V_{\rm end}}=e^{-3N_{\rm rh}(1+w_{\rm rh})}\,,
\label{eq:ratiorhend}
\ee
where we used $
\rho\propto a^{-3(1+w_{\rm rh})}$ during the reheating era. Also for the relation between $\rho_{\rm end}$ and $V_{\rm end}$, we used the fact that the kinetic energy is approximately half of the potential energy at the end of inflation defined by $\epsilon=1$. Eq.~(\ref{eq:ratiorhend}) allows one to express $T_{\rm rh}$ in terms of $N_{\rm rh}$ and $V_{\rm end}$, i.e.
\be
T_{\rm rh}=1.46\times\left(\frac{V_{\rm end}}{g_{*,\rm rh}}\right)^{1/4}e^{-3N_{\rm rh}(1+w_{\rm rh})/4}\,.
\label{eq:Trh}
\ee

Finally the substitution of Eq.~(\ref{eq:Trh}) and Eq.~(\ref{eq:Nrd}) into Eq.~(\ref{eq:scaleratio2}) yields the number of the e-foldings during the reheating era
\ba
N_{\rm rh}&=&\frac{4}{1-3w_{\rm rh}}\cr\cr
&\times&\left[-N_{\star}-\ln\left(\frac{k_{\star}}{a_{0}T_{0}}\right)+\ln\left(\frac{g_{*\rm rh}^{1/4}}{g_{*s,\rm  rh}^{1/3}}\right)-\frac{1}{4}\ln\left(V_{\rm end}\right)+\frac{1}{2}\ln\left(\frac{\pi^{2}rA_{s}}{2}\right)+7.5\times10^{-2}\right]\cr\cr
&\equiv&\frac{4}{1-3w_{\rm rh}}\times\left[-N_{\star}+N_{\rm upper}\right]\,,\nonumber\\
\label{eq:Nrh}
\ea
where $N_{\rm upper}$ is all but $-N_{\star}$ in the square bracket in Eq.~(\ref{eq:Nrh}). As promised, for a given $w_{\rm rh}$, we see that the prediction of an inflation model for $N_{\star}$, $r$, $A_{s}$ and $V_{\rm end}$ can determine $N_{\rm rh}$ in Eq.~(\ref{eq:Nrh}) and thus $T_{\rm rh}$ in Eq.~(\ref{eq:Trh}).
\subsection{Is $T_{\rm rh}\lesssim10^{9}{\rm GeV}$ consistent with the model?}
\label{sec:Trhconsistent}
As was pointed out in the last part of Sec.~\ref{sec:various}, as long as $T_{\rm rh}\lesssim10^{9}{\rm GeV}$ can be realized, the model can maintain the current minimal form without asking more fields either to generate gaugino mass through other mediation mechanisms than the anomaly mediation or to have an alternative DM candidate. If not ($T_{\rm rh}$ cannot be smaller than $10^{9}{\rm GeV}$), the model should be extended so as to have a new LSP and DM candidate other than the wino. In this section, we address this issue by computing the model's prediction for $T_{\rm rh}$ based on Sec.~\ref{sec:Trhandwrh}.

\begin{figure}
  \centering
  \includegraphics[width=0.7\hsize]{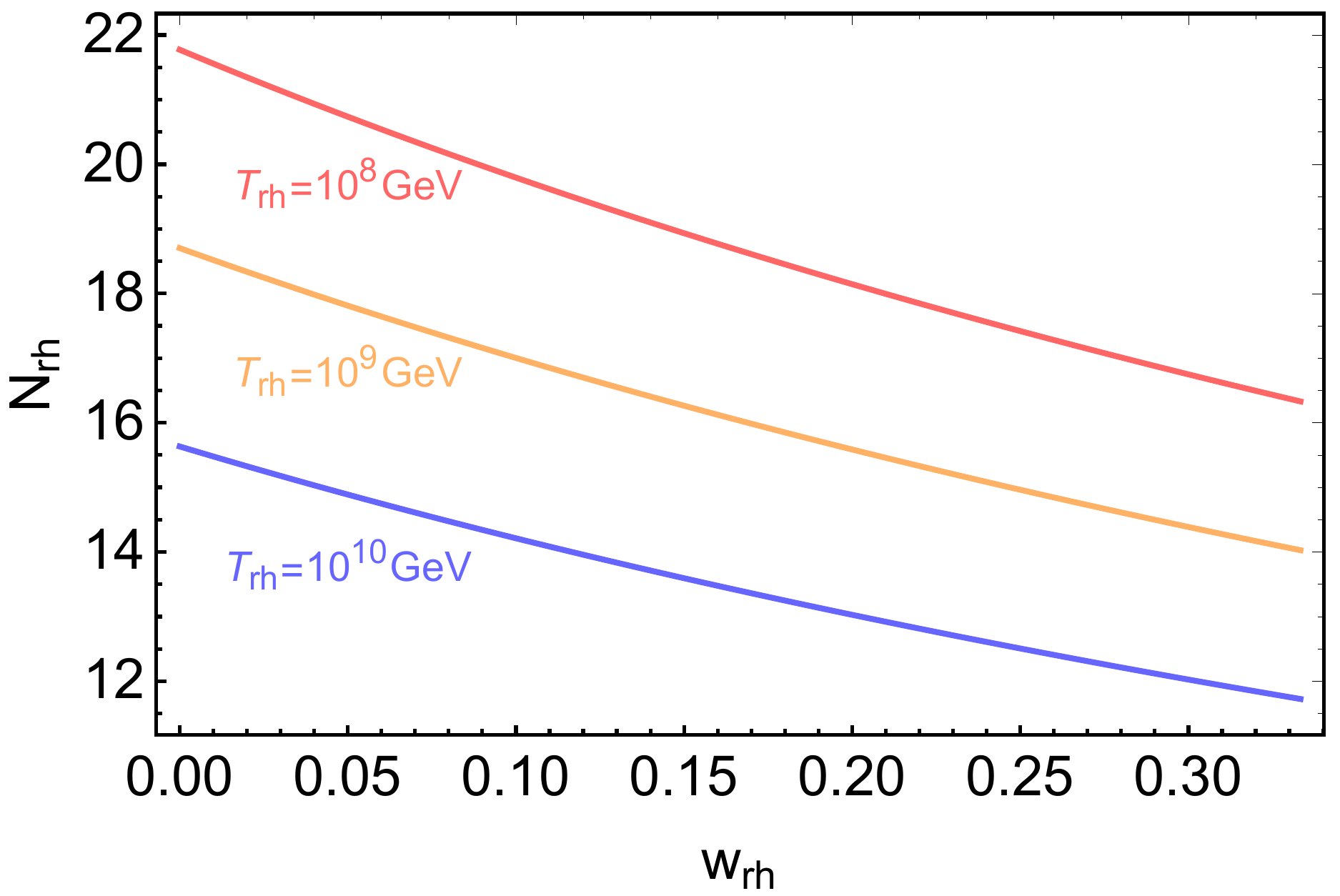}
  \caption{The relation between the equation of state of the universe during the reheating era ($w_{\rm rh}$) and the number of the e-foldings during reheating era for $g_{*,\rm rh}=230$ and $V_{\rm end}\simeq3\times10^{-12}$.}
  \label{fig2}
\end{figure}

For checking the consistency, we first attend to the relation between $N_{\rm rh}$ and $w_{\rm rh}$ in Eq.~(\ref{eq:Nrh}). For the inflaton potential explaining the CMB observables, we find $V_{\rm end}\simeq3\times10^{-12}$. Then substituting this $V_{\rm end}$ and $g_{*\rm rh}\simeq230$ in MSSM into Eq.~(\ref{eq:Nrh}), we obtain the relation between $N_{\rm rh}$ and $w_{\rm rh}$ for each $T_{\rm rh}$. This is shown in Fig.~\ref{fig2}. One can see that the smaller $T_{\rm rh}$ requires the larger $N_{\rm rh}$ for a fixed $w_{\rm rh}$. 

On the other hand, $N_{\rm upper}$ in Eq.~(\ref{eq:Nrh}) is approximately $53$ for $A_{s}=2.1\times10^{-9}$, $r=\mathcal{O}(10^{-4})$ and $V_{\rm end}\simeq3\times10^{-12}$. Because of the rapid coherent oscillation of the inflaton field with an approximately quadratic potential after the inflation ends, we expect that $w_{\rm rh}$ is close to $0$ for most of the time till the end of reheating. But considering details of the end of inflation and the process of reheating, we conservatively take $0\lesssim w_{\rm rh}\lesssim1/3$ and so $N_{\star}$ smaller than $\sim53$ is required for making $N_{\rm rh}$ positive in Eq.~(\ref{eq:Nrh}). 

Therefore, the inflation model in Sec.~\ref{sec:inflation} with $v\sim1.5\times10^{-3}$ can be consistent with $T_{\rm rh}$ as small as $10^{9}{\rm GeV}$ insofar as a  set of ($N_{\star},w_{\rm rh}$) produces a large enough $N_{\rm rh}$, i.e., at least $N_{\rm rh}\gtrsim14$. Observing Eq.~(\ref{eq:Nrh}) closely, one may think that having $T_{\rm rh}$ as small as what one desires in the model (or $N_{\rm rh}\gtrsim14$) is not difficult by requiring $w_{\rm rh}$ to be close to $1/3$. However, given that $w_{\rm rh}$ parametrizes the time-averaged value of the actual time-evolving equation of state of the universe during the reheating state, we expect $w_{\rm rh}$ to deviate from (smaller than) $1/3$. Also, we found that satisfying CMB observations alone (especially the constraint on $n_{s}$) already requires $N_{\star}\gtrsim42$. 

For these reasons, it is non-trivial to see whether $N_{\rm rh}$ allowing for $T_{\rm rh}$ as small as $10^{9}${\rm GeV} can be obtained while being consistent with all CMB observations. Given this question for the consistency, we go through the procedure to compute $T_{\rm rh}$ based on Sec.~\ref{sec:Trhandwrh}. And we confirmed that our inflation model characterized by $v=1.5\times10^{-3}$ and $r=\mathcal{O}(10^{-4})$ can indeed give rise to $T_{\rm rh}$ as small as $10^{9}{\rm GeV}$ for $0\lesssim w_{\rm rh}<1/3$ with $48.5\lesssim N_{\star}\lesssim53$, which is shown in the right panel of Fig.~\ref{fig1}.\footnote{Also the model is further specified by $g_{*,\rm rh}=g_{*s,\rm rh}\simeq230$, $c_{1}\gtrsim0.4$, $\kappa\sim0.3$, $\phi_{\star}\sim0.6$ and $\phi_{\rm end}\sim2$.} Especially for $w_{\rm rh}$ close to $0$, we found $N_{\star}=48.5-49$ is needed.

The larger $w_{\rm rh}$ makes it easier for the model to have the smaller $T_{\rm rh}$. Now that $T_{\rm rh}\lesssim10^{9}{\rm GeV}$ can be indeed realized, the gaugino mass in the model can be explained based on the anomaly mediation and the wino can be the DM candidate. In the next section, we study another way of probing $T_{\rm rh}$ based on the spectrum of GW sourced by the short-lived cosmic string present during the reheating era.

We end this section by pointing out a potential main channel for the inflaton decay for the case of $T_{\rm rh}\simeq10^{9}{\rm GeV}$. In the K\"{a}hler potential, we may expect the nonrenormalizable operator $\mathcal{O}_{\Phi N}=c_{\Phi N}|\Phi|^{2}|N|^{2}$ with $c_{\Phi N}=\mathcal{O}(1)$. Now that the decay rate of the inflaton due to the operator $\mathcal{O}_{\Phi N}$ reads $\Gamma(\phi\rightarrow2N)\simeq(m_{N}/M_{P})^{2}(m_{\Phi}/8\pi)\simeq1{\rm GeV}$, the comparison $\Gamma(\phi\rightarrow2N)\simeq H$ yields $T_{\rm rh}\simeq10^{9}{\rm GeV}$. Therefore, $T_{\rm rh}\simeq10^{9}{\rm GeV}$ can be understood in the perturbative reheating case thanks to the large enough $m_{N}=2c_{ij}\langle Q_{i}Q{j}\rangle$ (and thus $R[N]=0$). We notice that the consistency of the model with $(T_{\rm rh},m_{N})\simeq(10^{9}{\rm GeV},10^{11-12}{\rm GeV})$ is remarkable in the context of the seesaw mechanism for neutrino masses~\cite{Yanagida:1979as,Yanagida:1979gs,GellMann:1980vs,Minkowski:1977sc} and the primordial leptogenesis~\cite{Fukugita:1986hr,Buchmuller:2005eh}.

\section{Gravitational Wave: a potential smoking-gun feature in the future}
\label{sec:GW}
The scalar potential of a SUSY model is contributed by $F$-terms and $D$-terms. For a renormalizable scalar potential, there can be a direction in the space of complex scalars along which the potential vanishes as far as SUSY is respected. This direction is referred to as a ``flat direction" and the collection of such directions form the so-called ``moduli space". The flat directions, however, are lifted when soft masses for scalars are generated on SUSY-breaking.

Particularly for MSSM, before the SUSY-breaking takes place, there are many almost flat directions which can be conveniently characterized by gauge invariant monomials~\cite{Gherghetta:1995dv}. The correspondence between flat directions and gauge invariant monomials underlies this fact~\cite{Buccella:1982nx,Affleck:1983mk,Affleck:1984xz} and thereby the study of dynamics of a flat direction reduces to understanding gauge invariant operators and the scalar potential stemming from those. After the SUSY-breaking, flat directions are lifted since there appear unavoidable soft mass terms $m_{\rm soft}^{2}|\sigma|^{2}$ in the scalar potential where $\sigma$ collectively denotes the scalar components of chiral superfields.

Given that flat directions are so common in the SUSY theories, one may wonder if their dynamics can be used to test predictions of a SUSY model. Concerning this, the Hubble induced mass which the flat directions receive during the inflation and the reheating times could play a critical role. Suppose the sign of the coupling $|S|^{2}|\Sigma|^{2}$ is positive while that of the coupling $|\Phi|^{2}|\Sigma|^{2}$ is negative where $\Sigma$ is a flat direction and $S$ and $\Phi$ fields are defined in Eq.~(\ref{eq:WR2}). This gives rise to the situation where the sign of the Hubble induced mass is positive during the inflation while it is negative during the reheating stage.\footnote{The opposite situation is assumed in the Affleck-Dine baryogenesis scenario~\cite{Affleck:1984fy} and in the case where the primordial coherently oscillating scalar initiates the dark sector particles (see, for instance, \cite{Choi:2020nan,Choi:2020dec}).} Then a flat direction obtains a non-zero VEV after the inflation ends although it stays at the origin of the field space during the inflation. This implies that there can be formation of cosmic strings (CS) provided chiral superfields making up a gauge invariant monomial of interest carry a global $U(1)$ charge. If so, the gravitational wave (GW) generated by the CS can contain information for $m_{\rm soft}$. This is because the CSs are expected to disappear once the Hubble expansion rate during the reheating era becomes comparable to $m_{\rm soft}$. Namely, the time when the generation of GWs ceases is determined by $m_{\rm soft}$, which might be imprinted in the spectrum of the GW.

In \cite{Kamada:2014qja,Kamada:2015iga}, precisely this possibility was considered and it was confirmed via the simulation that the CS network forms and reaches the scaling regime prior to disappearance of the CSs. On top of that, it was also studied how $T_{\rm rh}$ and $m_{\rm soft}$ can be read from the spectrum of GW spectra generated by CSs. Now having the model featured by $m_{\rm soft}\simeq100-1000{\rm TeV}$ and $T_{\rm rh}\lesssim10^{9}{\rm GeV}$, in this section, we study how the prediction of the model can be imprinted in the GW possibly generated by the short-lived CSs originated from the temporary breaking of a global $U(1)$ symmetry. We will see that Einstein Telescope (ET) and Cosmic Explorers (CE) can be used to probe our scenario via GW detection.

\subsection{Flat Direction and Cosmic String Formation}
In this section, firstly, we point out the richness of flat directions which can obtain time independent VEVs determined by the K\"{a}hler potential rather than the super potential. After that, we compute the VEV of flat directions of our interest. The VEV lasts during the reheating era until the time when $H\simeq m_{\rm soft}$ is reached. We consider the situation wherein the spontaneous breaking of the global $U(1)_{B}$ (baryon charge) is induced by the VEV and also results in the formation of the global CSs. On disappearance of the VEV, CSs do as well and thus CSs are of the short-lived kind.

In our model, as was specified in Table.~\ref{table1}, R-charges of $H_{u}$, $H_{d}$ and $N$ are given by $R[H_{u}]+R[H_{d}]=4$ and $R[N]=0$. Along with these, the requirement that three Yukawa coupling operators have total R-charge 2 mod 6  can fully determine R-charges of the MSSM matter sector consisting of ${\bf 10}$, ${\bf \bar{5}}$, $H_{u}$, $H_{d}$ and $N$: $R[{\bf 10}]=0$, $R[{\bf \bar{5}}]=0$, $R[H_{u}]=2$, $R[H_{d}]=2$, $R[N]=0$.\footnote{For convenience, we use representations of $SU(5)_{\rm GUT}$ to refer to these field, i.e. ${\bf 10}=(q,\bar{u},\bar{e})$, ${\bf \bar{5}}=(\bar{d},\ell)$. Each field denotes quark $SU(2)_{L}$ doublet ($q$), lepton $SU(2)_{L}$ doublet ($\ell$), up-quark $SU(2)_{L}$ singlet ($\bar{u}$), down-quark $SU(2)_{L}$ singlet ($\bar{d}$), lepton $SU(2)_{L}$ singlet $\bar{e}$, and up and down type Higgs ($H_{u}$ and $H_{d}$).}

Given the concrete R-charge assignment, we encounter one remarkable consequence of the model concerning the contribution of a flat direction to the superpotential. That is, whenever the flat directions associated with gauge invariant monomials made of ${\bf 10}$ and ${\bf \bar{5}}$ appear in the superpotential, both renormalizable and nonrenormalizable operators of the flat directions must be accompanied by the suppression by the factor $(m_{3/2}/M_{P})\sim10^{-12}$. This is because the R-charge of an operator in the superpotential should be 2 mod 6.\footnote{Some operators in the superpotential are suppressed by $Z_{4}$ symmetry given in Table.~\ref{table1} since $Z_{4}$ charges are 1 for ${\bf 10}$ and ${\bf \bar{5}}$.} 

Let us refer to the flat direction associated with gauge invariant monomials made of ${\bf 10}$ and ${\bf \bar{5}}$ as $\chi$. In Table.~3 of \cite{Gherghetta:1995dv}, one can find gauge invariant monomials in MSSM which can be used to write any gauge invariant polynomial in $(q,\ell,\bar{u},\bar{d},\bar{e},H_{u},H_{d})$. As an exemplary operator, we may attend to $\bar{u}\bar{d}\bar{d}$. $\chi$ being as the flat direction of $\bar{u}\bar{d}\bar{d}$, its superpotential is given by
\be
W_{\chi}=\left(\frac{m_{3/2}}{M_{P}}\right)\sum_{p=3}a_{\chi,p}\frac{\chi^{p}}{M_{P}^{p-3}}\quad\rightarrow\quad V(\chi)\ni \left(\frac{m_{3/2}}{M_{P}}\right)^{2}\sum_{p=3}a_{\chi,p}^{2}\frac{\chi^{2p-2}}{M_{P}^{2p-6}}\qquad(p\in\mathbb{Z}) \,,
\label{eq:Wchi}
\ee
where $R[m_{3/2}]=2$ and $R[\chi^{p}]=0$, and $a_{\chi,p}$ is a dimensionless coefficient. Here we wrote $M_{P}$ explicitly for the clarity. When compared to a term of the same mass dimension from K\"{a}hler potential, due to $m_{3/2}<<H$ before CSs disappear (see Eq.~(\ref{eq:Vrh})), contributions to $V(\chi)$ in Eq.~(\ref{eq:Wchi}) are negligible for determining the VEV of $\chi$. Hence, it is K\"{a}hler potential that determines the VEV of $\chi$ in our model.

This result applies not only to the flat direction of $\bar{u}\bar{u}\bar{d}$ but to any flat direction associated with gauge invariant monomials purely composed of ${\bf 10}$ and ${\bf \bar{5}}$. And this ensures the richness of flat directions which can potentially satisfy conditions for the signs of the Hubble induced masses we require. This unavoidable feature works in model's favor in terms of the strength of GW signal induced by the short-lived CSs. In \cite{Kamada:2015iga}, the resultant GW spectra were studied for each of the cases differentiated by which one determines the VEV of a flat direction among superpotential or K\"{a}hler potential. It turns out that the information for $m_{\rm soft}$ and $T_{\rm rh}$ can be imprinted in GW spectra equally for both cases. However, the strength of the GW spectra is relatively larger when K\"{a}hler potential determines the VEV of a flat direction. This fact renders our model more advantageous in justifying the higher chance of producing the larger GW signal in comparison with other SUSY models. Again this is essentially attributable to the assumed discrete gauged $Z_{6R}$ symmetry.

With the assumed positive Hubble induced mass and also $m_{\rm soft}$ generated at the end of the inflation, $\chi$ is expected to sit in the origin of the field space during inflation. After inflation ends, the reheating era gets started and we consider the following K\"{a}hler potential of $\chi$
\be
K\supset\frac{a_{2}}{M_{P}^{2}}|\Phi|^{2}|\chi|^{2}\,+\,\frac{a_{n}}{M_{P}^{2n-2}}|\Phi|^{2}|\chi|^{2n-2}\,,
\label{eq:Krh}
\ee
where $a_{2}$ ($a_{n}$) are dimensionless coefficients of operators of mass dimension $4$ ($2n$), and $n$ is a positive integer greater than 2. After integrating over the superspace coordinates, there arise terms including $|\dot{\phi}|^{2}$ which result in the following potential for $\chi$
\be
V(\chi)=\left[\frac{3a_{2}}{2}H^{2}+m_{\rm soft}^{2}\right]|\chi|^{2}+\frac{3a_{n}}{2}H^{2}\frac{|\chi|^{2n-2}}{M_{P}^{2n-4}}\,,
\label{eq:Vrh}
\ee
where we used the equipartition of the energy density of $\phi$ during oscillation, i.e. $|\dot{\phi}|^{2}/2=\rho_{\phi}/2\simeq(3/2)H^{2}M_{P}^{2}$. Given $H_{\rm inf}\simeq v^{2}\simeq10^{12}{\rm GeV}$ and $m_{\rm soft}\simeq100-1000{\rm TeV}$, it can be seen easily that the Hubble induced mass dominates over the soft mass term from the end of inflation to the time when $H\simeq m_{\rm soft}$ is reached. If $a_{2}<0$ and $a_{n}>0$ hold, the flat direction obtains the non-vanishing VEV
\be
\langle|\chi|\rangle=\left(\frac{|a_{2}|}{a_{n}(n-1)}\right)^{\frac{1}{2n-4}}M_{P}\,.
\label{eq:chivev}
\ee
Note that this VEV is independent of time. Once this VEV is acquired by the flat direction which is a linear combination of squarks or sleptons, $U(1)_{\rm B}$ symmetry becomes spontaneously broken and this temporary breaking lasts until the time when the two terms in the square bracket in Eq.~(\ref{eq:Vrh}) are comparable is reached. At this time ($t=t_{\rm decay}$), the cosmic string starts to decay as $V(\chi)$ becomes the positive curvature potential. Accordingly, $\chi$ starts to oscillate around $\chi=0$ and eventually sits at the origin.\footnote{In our work, we focus on flat directions including 3rd generation quark fields like $\bar{u}_{3}\bar{d}_{2}\bar{d}_{3}$ where the subscripts are generation indices. This makes the one-loop correction to the scalar potential dominated by that due to Yukawa interaction. In this case, the scalar potential is steeper than the quadratic one~\cite{Enqvist:1997si}, preventing the B-ball formation after CSs disappear.   }

\subsection{GW Spectrum and Testing the Model}
The current spectrum of the GW ($\Omega_{\rm GW}h^{2}(f,t_{0})$) induced by the CS that exists since the end of the inflation ($t=t_{\rm end}$) until the time of $H\simeq m_{\rm soft}$ ($t=t_{\rm decay}$) is characterized by three ranges of Fourier modes: modes entering the horizon at (i) $t<t_{\rm decay}$, (ii) $t_{\rm decay}<t<t_{\rm rh}$ and (iii) $t>t_{\rm rh}$ where $t_{\rm rh}$ is the time when the temperature of the universe reaches $T_{\rm rh}$. These three regimes are distinguished by $f_{\rm peak}$ and $f_{\rm rh}$. The former (later) is the GW frequency today corresponding to the $k$-mode that re-entered the horizon at the time when CSs decays (when $T\simeq T_{\rm rh}$ holds).\footnote{Note that for $T_{\rm rh}$ and $m_{\rm soft}$ of our interest, $T_{\rm max}\simeq0.5\times T_{\rm rh}^{1/2}H_{\rm inf}^{1/4}M
_{P}^{1/4}$~\cite{Kolb:1990vq,Chung:1998rq} is larger than the temperature when $H\simeq m_{\rm soft}$ holds. Thus, there can be indeed the time interval when CSs form and exist prior to their decay.}

For the first regime of $k$ re-entering the horizon at the time when the CSs form, reach the scaling regime and decay ($k\gtrsim k_{\rm peak}$), the time evolution of the GW energy density was studied in \cite{Kamada:2015iga} by solving the time evolution equation of $\chi$ numerically.\footnote{In \cite{Kamada:2015iga}, the oscillation domination was assumed, i.e. $H\propto a^{-3/2}$ and $w_{\rm rh}=0$, in performing the lattice simulation for solving the time evolution equation of $\chi$. Nevertheless, the presence of the scaling regime is expected not to be affected even for $w_{\rm rh}$ other than 0 as long as $w_{\rm rh}>-1/3$. We are grateful to M. Yamada for pointing out this.} With $\Omega_{\rm GW}$ defined in Eq.~(\ref{eq:OmegaGW}), ${\rm d}\Omega_{\rm GW}/{\rm d}\log\tau$ was found to have a peak (GW energy production is most efficient) at $k$ of which size is equal to $\sim40\%$ of the comoving Hubble radius. Because of that, $\Omega_{\rm GW}(\tau)$ has a peak which keeps shifting to the smaller $k$ (larger length scale) with time until $t_{\rm decay}$ is reached. Since then, the comoving $k$-mode at the peak ($k_{\rm peak}$) is frozen and the redshifted peak structure remains to date. 

The GW energy density at $k_{\rm peak}$ today reads~\cite{Kamada:2015iga}
\be
\Omega_{\rm GW}^{\rm peak}h^{2}\simeq5\times10^{-9}\left(\frac{|a_{2}|^{-1/2}m_{\rm soft}}{10^{3}{\rm TeV}}\right)^{-2/3}\left(\frac{T_{\rm rh}}{10^{9}{\rm GeV}}\right)^{4/3}\left(\frac{|a_{2}|}{a_{n}(n-1)}\right)^{\frac{2}{n-2}}\,,
\label{eq:GWpeak}
\ee
and the corresponding peak frequency today is
\be
f_{\rm peak}\simeq7000{\rm Hz}\left(\frac{|a_{2}|^{-1/2}m_{\rm soft}}{10^{3}{\rm TeV}}\right)^{1/3}\left(\frac{T_{\rm rh}}{10^{9}{\rm GeV}}\right)^{1/3}\,.
\label{eq:fpeak}
\ee

Next, for the other regime of $k$ re-entering the horizon at the time $t>t_{\rm decay}$, the GW spectrum can be obtained from Eq.~(\ref{eq:finalspectrum}) with the numerically computed Eq.~(\ref{eq:Aij}) and (\ref{eq:Bij}). Before reheating completes, the equation of state of the universe is $w_{\rm rh}$ while it is $1/3$ after the reheating completes. This means that particularly for $w_{\rm rh}$ as small as $0$, the $k$-dependence of $\Omega_{\rm GW}$ for $k<k_{\rm rh}$ and $k\gtrsim k_{\rm rh}$ is expected to be clearly distinguishable~\cite{Seto:2003kc,Nakayama:2008ip} due to different dilution of $\Omega_{\rm GW}$. In \cite{Kamada:2015iga}, $w_{\rm rh}\simeq0$ case was studied.

For our model, as we discussed in Sec.~\ref{sec:Trhandwrh}, in principle any value lying in $0\lesssim w_{\rm rh}<1/3$ can be possible but $w_{\rm rh}$ being close to $0$ is more realistic. For the purpose of the potential clear bending signature of $\Omega_{\rm GW}$, from here on we focus on the case with $w_{\rm rh}\simeq0$ but with $T_{\rm RH}=10^{8}-10^{9}{\rm GeV}$.

\begin{figure}
\centering
  \includegraphics[scale=0.48]{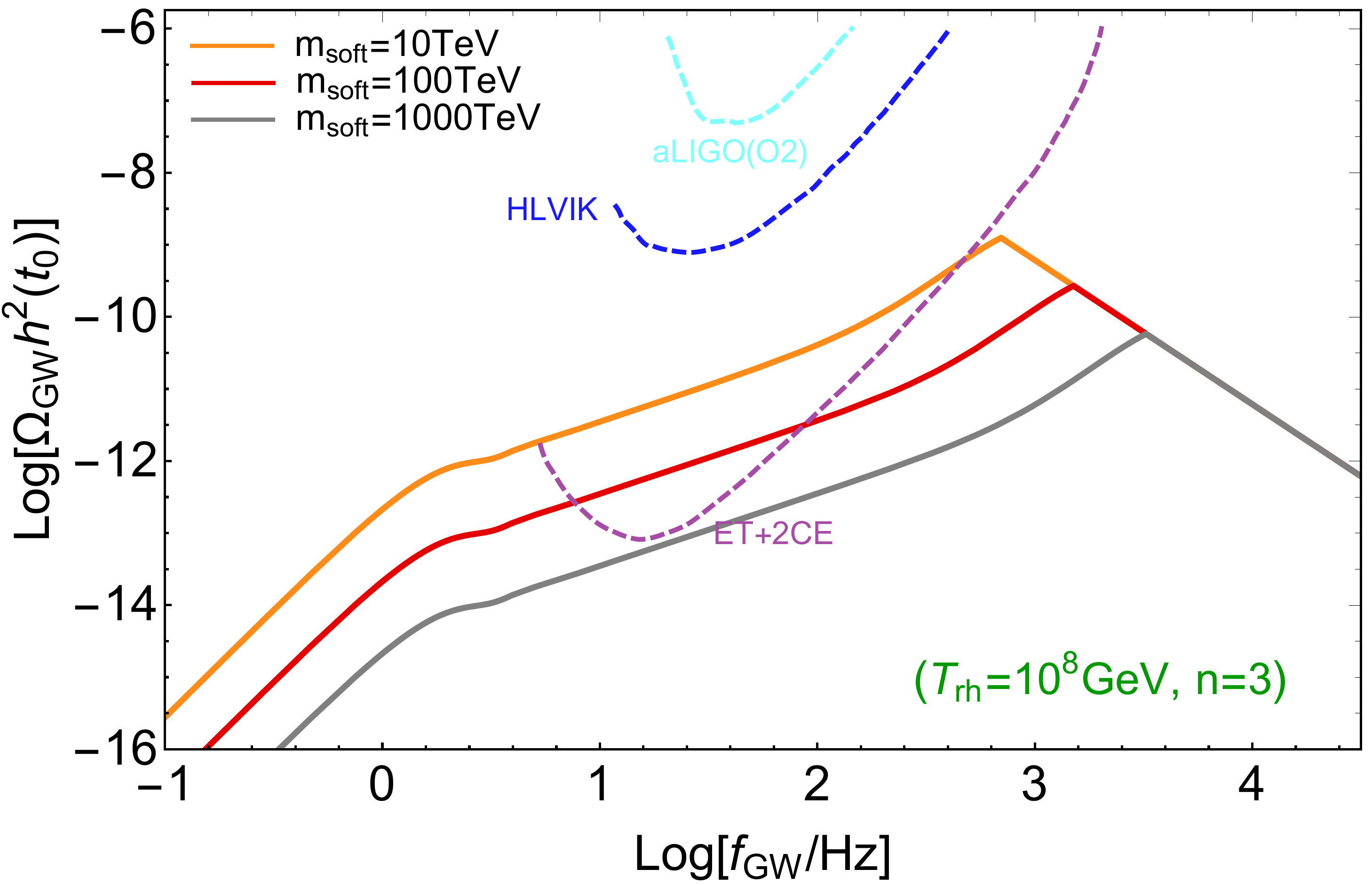}
  \includegraphics[scale=0.48]{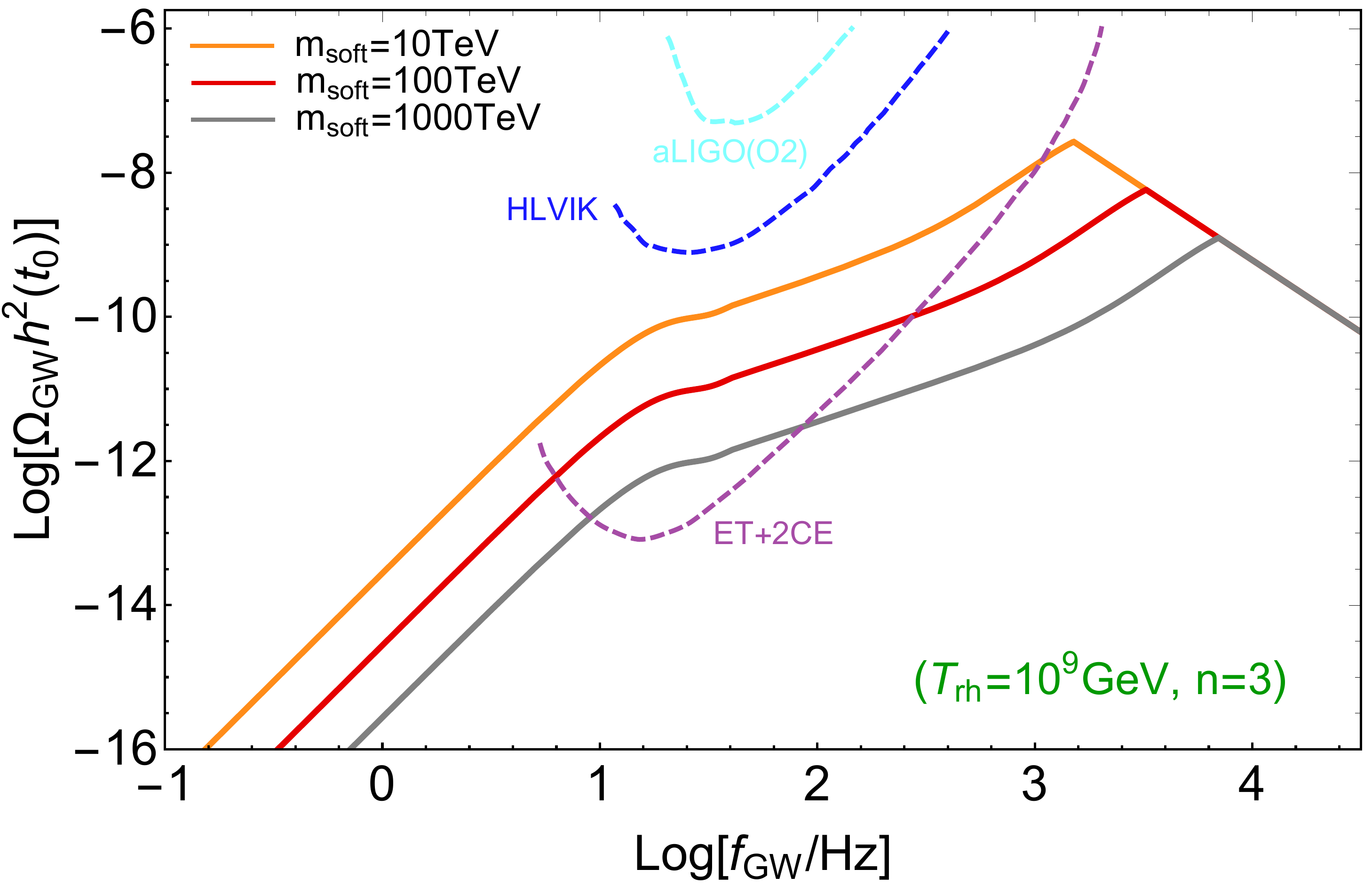}
  \caption{GW spectra for different $T_{\rm rh}$ and $m_{\rm soft}$. For both panels, $a_{2}=a_{n}=1$ and $n=3$ are commonly assumed. Also each of the solid lines with yellow, red and gray colors corresponds to $m_{\rm soft}=10$, $100$, and $1000{\rm TeV}$. The upper (lower) panel shows the case with $T_{\rm rh}=10^{8}{\rm GeV}$ ($10^{9}{\rm GeV}$).}
  \label{fig3}
\end{figure}

Based on $f_{\rm GW}=k/(2\pi a_{0})$ and the entropy conservation, the GW frequency today corresponding to $k_{\rm rh}$ reads
\be
f_{\rm rh}=\left(\frac{g_{s}(t_{0})}{g_
{s}(t_{\rm rh})}\right)^{1/3}\left(\frac{T_{0}}{T_{\rm rh}}\right)\frac{k_{\rm rh}}{2\pi a_{\rm rh}}\simeq30{\rm Hz}\left(\frac{T_{\rm rh}}{10^{9}{\rm GeV}}\right)\,,
\label{eq:frh}
\ee
where we used $k_{\rm rh}=a_{\rm rh}H(a_{\rm rh})$ for the second equality. At the frequency in Eq.~(\ref{eq:frh}), $\Omega_{\rm GW}(k)$ is expected to reveal the bending due to different $k$-dependence ascribable to the change in the equation of state of the universe before and after the reheating.

In Fig.~\ref{fig3}, we show the GW spectra corresponding to various different cases of $(T_{\rm rh},m_{\rm soft})$. For $f<f_{\rm peak}$, we numerically compute $\Omega_{\rm GW}h^{2}$ in accordance with Appendix.~\ref{sec:appendixA} with the constant $T_{ij}^{TT}$.\footnote{The modes satisfying $k<k_{\rm peak}$ were outside of the horizon when GW was generated by CSs. For those superhorizon modes, the lack of causality makes $T_{ij}^{TT}$ independent of $k$.} For $f>f_{\rm peak}$, we can obtain $\Omega_{\rm GW}h^{2}$ based on the fact that $\Omega_{\rm GW}\propto f^{-2}$~\cite{Kamada:2015iga} and the use of Eq.~(\ref{eq:GWpeak}) and (\ref{eq:fpeak}). For both panels, $a_{2}=a_{n}=1$ and $n=3$ are commonly assumed. Solid lines show the GW spectra whereas the dashed lines are the sensitivity curves of upcoming GW experiments. Solid lines of different colors correspond to the specified $m_{\rm soft}$=$10$TeV (yellow), $100$TeV (red), $1000$TeV (gray). The sensitivity curve (purple dashed) of Einstein Telescope (ET)~\cite{Punturo:2010zz} and two third generation Cosmic Explorers (CE)~\cite{Reitze:2019iox} is read from \cite{Perigois:2020ymr}. We also shows the sensitivity curves of the Advanced LIGO O2 (cyan dashed)~\cite{LIGOScientific:2019vic} and HLVIK (blue dashed)~\cite{LIGOScientific:2014pky,LIGOScientific:2013rhu,Aso:2013eba,Unnikrishnan:2013qwa}.\footnote{HLVIK is the network of several terrestrial GW detectors including Advanced LIGO Hanford, and Livingston, Advanced Virgo, LIGO India, and KAGRA.} The upper (lower) panel shows the case with $T_{\rm rh}=10^{8}{\rm GeV}$ ($10^{9}{\rm GeV}$).

We see that ET+2CE may have a chance to see the GW spectrum induced by the short-lived cosmic strings provided $T_{\rm rh}$ is as large as $10^{8}-10^{9}{\rm GeV}$ and $m_{\rm soft}=10-1000{\rm TeV}$. Particularly, for $T_{\rm rh}\simeq10^{9}{\rm GeV}$, the bending at $f_{\rm rh}$ can be seen by ET+2CE, which can tell us the value of $T_{\rm rh}$ directly via Eq.~(\ref{eq:frh}). Ideally, for the case where both $\Omega_{\rm GW}^{\rm peak}h^{2}$ and $\Omega_{\rm GW}^{\rm bend}h^{2}$ are within the sensitivity curve at $f=f_{\rm peak}$ and $f=f_{\rm rh}$ respectively, $m_{\rm soft}$ and $T_{\rm rh}$ can be directly read from Eq.~(\ref{eq:GWpeak}), Eq.~(\ref{eq:fpeak}) and Eq.~(\ref{eq:frh}). However, we can see that the bending at $f_{\rm rh}$ is easier to 
observe, e.g., for $T_{\rm rh}=10^9{\rm{GeV}}$. We find to a good approximation that $\frac{\Omega^{\rm bend}_{\rm GW}}{\Omega^{\rm peak}_{\rm GW}}\simeq0.365\frac{f_{\rm rh}}{f_{\rm peak}}$ and so,
\be
\Omega^{\rm bend}_{\rm GW}h^2\simeq7.8\times10^{-12}\left(\frac{|a_{2}|^{-1/2}m_{\rm soft}}{10^{3}{\rm TeV}}\right)^{-1}\left(\frac{T_{\rm rh}}{10^{9}{\rm GeV}}\right)^2\left(\frac{|a_{2}|}{a_{n}(n-1)}\right)^{\frac{2}{n-2}}\,.
\label{eq:GWbend}
\ee
Thus, even if we can only resolve the bending in the GW spectrum, we will be able to infer both $T_{\rm rh}$ and $m_{\rm soft}$ with Eq.~\eqref{eq:frh} and Eq.~\eqref{eq:GWbend}. 

It is worth pointing out that the feature that the peak and bending locations correlate with their amplitudes in such ways is difficult to be realized in other models. While such a GW spectrum is not a necessary prediction, it would be a smoking gun for our model once the two characteristic frequencies are observed in the future.

\section{Conclusions}
Discrete R-symmetry $Z_{NR}$ ($N\in\mathbb{Z})$ is a very interesting possibility in SUSY models in that its anomaly free conditions can be satisfied within MSSM. Particularly $Z_{6R}$ is special in that it is the unique $Z_{NR}$ that is free of the mixed-anomalies for three generations of quarks and leptons in MSSM. Should the discrete R-symmetry plays an important role in low energy physics, however, the relevant domain wall problem becomes a severe issue since discrete symmetries are most probably gauged~\cite{Krauss:1988zc}.

On the other hand, $R$-symmetry is analogous to spacetime symmetries in that every operator in the superpotential has to respect the $R$-symmetry. This fact may indicate an interesting possibility that some of dimensionful parameters in SUSY models can be powers of a $R$-symmetry breaking scale. Put it another way, knowing that $R$-symmetry must be broken in SUGRA for having a constant term in the superpotential and the breaking should be induced by a field with a non-zero R-charge, we may imagine the situation in which some dimensionful parameters are nothing but spurions of the broken $R$-symmetry in low energy physics.

Motivated by these points, in this work, we considered the possibility in which the gauged $Z_{6R}$ is spontaneously broken by the formation of the condensation $\langle QQ\rangle$ in the confinement of the hidden strong dynamics of $Sp(2)$ prior to the inflationary era. The breaking at the energy scale $\sqrt{\langle QQ\rangle}\simeq v\simeq1.5\times10^{3}$ in turn drives the new inflation type potential with the VEV of the inflaton $\langle\Phi\rangle\simeq2$. With the non-zero $R[Q]=+1$, powers of the $Sp(2)$ invariant $QQ$ couple to $H_{u}H_{d}$ and $NN$ so that the confinement of the hidden strong dynamics of $Sp(2)$ also generates Higgsino mass $\mu_{H}\sim\langle QQ\rangle^{2}\sim v^{4}$ and the right-handed neutrino masses $m_{N}\sim\langle QQ\rangle\sim v^{2}$. Embedded in SUGRA framework, the model predicts the scalar soft masses of order $m_{3/2}$ and EWSB further requires $\mu_{H}=\mathcal{O}(m_{3/2})$. Therefore, the model accounts for five energy scales for the inflation, the R-symmetry breaking, the SUSY-breaking, the Higgsino mass and the right-handed neutrino mass based on the common single origin, i.e. spontaneous R-symmetry breaking before inflationary era. This result is summarized in Table.~\ref{table1}.

The model being along the same line as the pure gravity mediation scenario~\cite{Ibe:2011aa}, it has wino as the DM candidate in its minimal form. For avoiding the overclosure of the universe due to too much abundance of wino DM, $T_{\rm rh}\lesssim10^{9}{\rm GeV}$ is required. We confirmed that the model can indeed lead to $T_{\rm rh}$ as small as $10^{9}{\rm GeV}$ as far as $w_{\rm rh}$ can be close to zero (see Sec.~\ref{sec:reheating}).

Finally, in Sec.~\ref{sec:GW}, we discussed the GW spectra induced by the short-lived CSs which can be a potential smoking gun experimental signal of the model. As is the case for other SUSY models, there can be many flat directions in the model characterized by gauge invariant monomials. If the flat direction associated with a gauge invariant monomial made of squark and slepton fields couples to $S$ and $\Phi$ with a positive and a negative coupling constant respectively in the K\"{a}hler potential, there can be temporary CSs that are present since the end of inflation until the time when $T\simeq T_{\rm rh}$ is satisfied. If this is the case, the information for $m_{\rm soft}$ and $T_{\rm rh}$ can be imprinted in the GW spectra caused by the shorted-lived CSs~\cite{Kamada:2014qja,Kamada:2015iga}. The noticeable consequence of the model is that the VEV of the flat direction is determined by the K\"{a}hler potential. This guarantees the strength of the GW spectra large enough to be detected by upcoming GW experiments including Einstein Telescope (ET) and Cosmic Explorers (CE) (see Fig.~\ref{fig3}). 

\section*{Acknowledgments}
G.C. is grateful to M. Yamada for the discussion about GW induced by the short-lived CSs. T. T. Y. is supported in part by the China Grant for Talent Scientific Start-Up Project and by Natural Science Foundation of China under grant No. 12175134 as well as by World Premier International Research Center Initiative (WPI Initiative), MEXT, Japan.

\appendix
\section{Computation for the GW Spectrum}
\label{sec:appendixA}
\setcounter{equation}{0}
In this section, we make a review of the way to compute the spectrum of the GW sourced by the shorted-lived cosmic string based on \cite{Dufaux:2007pt,Kawasaki:2011vv,Kamada:2015iga}. For more details, we refer the readers to \cite{Kamada:2015iga}.

In the Friedmann-Robertson-Walker (FRW) background, the GW is the traceless ($h^{i}_{i}=0$) and transverse ($\partial^{i}h_{ij}=0$) tensor fluctuation $h_{ij}$ as can be seen in
\be
ds^{2}=a(\tau)^{2}[-d\tau^{2}+(\delta_{ij}+h_{ij})dx^{i}dx^{j}]\quad(i,j=1,2,3)\,.
\label{eq:metric}
\ee
$h_{ij}(t,\bold{x})$ can be Fourier-expanded as
\be
h_{ij}(t,\bold{x})=\int\frac{d^{3}k}{(2\pi)^{3/2}}h_{ij}(t,\bold{k})e^{i\bold{k}\cdot\bold{x}}\,,
\ee
where $\bold{x}$ ($\bold{k}$) is the three position (momentum) vector. 

The GW spectrum as a function of the conformal time $\tau$ and the comoving wavenumber $k$ is defined to be
\be
\Omega_{\rm GW}(k,\tau)\equiv\frac{1}{\rho_{\rm total}(\tau)}\frac{{\rm d}\rho_{\rm GW}(k,\tau)}{{\rm d}\log k}\,,
\label{eq:OmegaGW}
\ee
where $\rho_{\rm total}(\tau)=3M_{P}^{2}H(\tau)^{2}$ the total energy density of the universe at the conformal time $\tau$ and $\rho_{\rm GW}(k,\tau)$ is the energy density of GW. The GW energy density is given by 
\be
\rho_{\rm GW}=\frac{1}{32\pi G}\langle\dot{h}_{ij}\dot{h}^{ij}\rangle_{V}=\frac{1}{32\pi G}\frac{\langle h_{ij}^{'}h^{ij\,'}\rangle_{V}}{a^{2}}\,,
\label{eq:rhoGW}
\ee
where the dot (prime) is the derivative with respect to the time $t$ (conformal time $\tau$). Here $\langle..\rangle_{V}$ means the average over a volume of the size of several wavelengths. 

In the presence of (traceless and transverse) anisotropic stress $T_{ij}^{TT}$ which lasts for the time inverval $[\tau_{i},\tau_{f}]$, the time evolution of the Fourier component of $h_{ij}$ is given by
\be
h_{ij}^{''}+2\frac{a^{'}}{a}h_{ij}^{'}+k^{2}h_{ij}=16\pi GT_{ij}^{TT}\,,
\label{eq:hijeqwithTij}
\ee
where $G\equiv(8\pi M_{P}^{2})^{-1}$ is the Newtonian constant. Having $h_{ij}(\tau_{i})=h_{ij}^{'}(\tau_{i})=0$ as the initial condition, the solution to Eq.~(\ref{eq:hijeqwithTij}) for $\tau\in[\tau_{i},\tau_{f}]$ can be obtained by the time integral of $T_{ij}^{TT}$ convoluted with a Green function~\cite{Dufaux:2007pt,Kawasaki:2011vv}. For the time $\tau>\tau_{\rm RH}>\tau_{f}=\tau_{\rm decay}$, $h_{ij}$ follows the time evolution equation without $T_{ij}^{TT}$ in Eq.~(\ref{eq:hijeqwithTij}) and the solution thereof is given by~\cite{Kamada:2015iga}
\be
h_{ij}(\bold{k},\tau)=A_{ij}(\bold{k})\frac{k\tau}{a}j_{0}(k\tau)+B_{ij}(\bold{k})\frac{k\tau}{a}n_{0}(k\tau)\,,
\label{eq:hijsol}
\ee
where the time independent coefficients $A_{ij}(\bold{k})$ and $B_{ij}(\bold{k})$ contain information for $T_{ij}^{TT}$ through
\be
A_{ij}(\bold{k})=-16\pi G\int_{\tau_{i}}^{\tau_{f}}d\tau\, \tau a(\tau)f_{A}(k\tau)T_{ij}^{TT}(\bold{k},\tau)\,,
\label{eq:Aij}
\ee
\be
B_{ij}(\bold{k})=16\pi G\int_{\tau_{i}}^{\tau_{f}}d\tau\, \tau a(\tau)f_{B}(k\tau)T_{ij}^{TT}(\bold{k},\tau)\,.
\label{eq:Bij}
\ee
with the following forms of $f_{A}(k\tau)$ and $f_{B}(k\tau)$
\be
f_{A}(k\tau)=a_{1}n_{1}(k\tau)-a_{2}j_{1}(k\tau)\,,
\label{eq:fA}
\ee
\be
f_{B}(k\tau)=-b_{1}n_{1}(k\tau)+b_{2}j_{1}(k\tau)\,,
\label{eq:fB}
\ee
where $j_{1}$ and $n_{1}$ are the spherical Bessel and Neumann function of the first order respectively. The coefficients $a_{1}$, $a_{2}$, $b_{1}$ and $b_{2}$ are given by
\be
a_{1}=x^{2}[j_{1}(x)\partial_{x}n_{0}(x)-n_{0}(x)\partial_{x}j_{1}(x)]\,,
\label{eq:a1}
\ee
\be
a_{2}=x^{2}[n_{1}(x)\partial_{x}n_{0}(x)-n_{0}(x)\partial_{x}n_{1}(x)]\,,
\label{eq:a2}
\ee
\be
b_{1}=x^{2}[j_{1}(x)\partial_{x}j_{0}(x)-j_{0}(x)\partial_{x}j_{1}(x)]\,,
\label{eq:b1}
\ee
\be
b_{2}=x^{2}[n_{1}(x)\partial_{x}j_{0}(x)-j_{0}(x)\partial_{x}n_{1}(x)]\,,
\label{eq:b2}
\ee
where $x$ is to be evaluated at $x=k\tau_{\rm RH}$. 
Finally, after substituting $h_{ij}$ in Eq.~(\ref{eq:hijsol}) into Eq.~(\ref{eq:rhoGW}) and Eq.~(\ref{eq:OmegaGW}), one obtains the spectrum of the GW
\be
\Omega_{\rm GW}\simeq\frac{k^{5}}{48\pi^{2}Va^{4}H^{2}}\sum_{ij}(|A_{ij}|^{2}+|B_{ij}|^{2})\,.
\label{eq:finalspectrum}
\ee

\bibliography{main}
\bibliographystyle{jhep}

\end{document}